\documentclass[baaa]{baaa}

\usepackage[pdftex]{hyperref}
\usepackage{subfigure}
\usepackage{natbib}
\usepackage{helvet,soul}
\usepackage[font=small]{caption}

\contriblanguage{1}

\contribtype{4}

\thematicarea{1}

\title{Spatially-resolved Evolution of Galaxies}
\subtitle{ }

\titlerunning{ }

\author{J.K. Barrera-Ballesteros\inst{1}}
\authorrunning{Barrera-Ballesteros J.K.}

\contact{jkbarrerab@astro.unam.mx}

\institute{
 Instituto de Astronom\'{i}a, Universidad Nacional Aut\'{o}noma de M\'{e}xico, A.P. 70-264, 04510 M\'{e}xico, D.F.,  Mexico
}

\resumen{}

\abstract{
Projected in the sky, galaxies are spatially-resolved objects. To understand how they formed and evolve it is necessary to study the spatial distribution of their observables. In this review talk, we briefly describe some scaling relations used to understand the physical processes that drive galaxy evolution, in particular for disk-like star-forming galaxies. First, we explore the relations derived using integrated galactic properties, then we introduce the scaling relations at kpc scales derived using the technique called Integral Field Spectroscopy (IFS) for large samples of galaxies in the nearby Universe.The very existence of scaling relations at kpc scales is a strong evidence that any physical scenario that explains the observed global scaling relations must be able to also explain their local counterpart. 
}


\begin{document}

\maketitle

\section{Introduction}
\label{sec:intro}

One of the most powerful tools that astronomers have to understand how galaxies form and evolve is through the so called scaling relations. These relations link different observational properties, depending on how tight they are, what dynamical range they cover, and their similarity across cosmic times, they give crucial information about the scenarios galaxies form and evolve.

Historically, 
one of the first studies that describe the spatial distribution of light in galaxies was the morphological classification of galaxies by E. Hubble \citep[][]{Hubble_1926}. In this scheme, galaxies are classified basically in two large families of morphologies: Redish or yellowish spheroidals 
and bluish disk-like galaxies with a wide variety of structures. 
In the last decade, with the use of IFS observational technique in extensive samples of galaxies it has been possible to determine the relation between observables at kpc-scales. In the next section (Sec.~\ref{sec:Global}) we will describe the main scaling relations for star-forming galaxies. This is a fundamental step to understand the scaling relations for star-forming regions derive at kpc scales (Sec. \ref{sec:Local}). A recent detailed review of scaling relations at kpc scales is presented by \cite{Sanchez_2019_rev}.  


To display both the global and scaling relations we will use along this article the dataset from the Sloan Digital Sky Survey SDSS-IV MaNGA survey \citep[Mapping Nearby Galaxies at Apache Point, ][]{Bundy_2015} in its 9th data release. This sample consists of more than 8000 galaxies located in the nearby Universe ($z > 0.15$) covering a wide range of stellar masses and morphologies. Therefore this sample can be considered as representative of the local Universe. The analysis of this dataset was performed by the analysis pipeline \textsc{Pipe3D} \citep{Sanchez_2016}.
\section{Scaling Relations for Global Galactic Properties}
\label{sec:Global}
Prior to acquisition of large datasets from spectroscopic surveys, photometric studies already revealed important information of the physical properties of galaxies. 
With large samples of galaxies it was evident that galaxies follow a clear relation in a color-magnitude diagram (CMD). Red galaxies 
follow a clear correlation with their luminosity. High-luminosity galaxies tend to be redder than low-luminosity ones following a tight correlation (known as the 'Red Sequence of Galaxies'). Whereas blue galaxies are located in the so-called 'Blue Cloud'.  
More than a mere classification tool, the CMD has been thought as an \textit{evolutionary plane} where young galaxies in the Blue Cloud evolve moving from this cloud to a intermediate zone called the 'Green Valley'. When aging, galaxies end up in the red-sequence. What are the exact physical processes that lead to this possible evolutionary path and what are the timescales that they operate are still in a large extend open questions for which astronomers are still looking for comprehensive answer. 

With the appearance of large spectroscopic surveys such as the Sloan Digital Sky Survey (SDSS), it was possible to determine scaling relations in large sample of galaxies using other observables.
In this section we will describe three fundamental scaling relations derived using integrated properties for star forming galaxies: The star-formation law (Sec.~\ref{sec:KS}), the star-formation main sequence (Sec.~\ref{sec:SFMS}), and the mass-metallicity relations (Sec.~\ref{sec:MZR}).

\subsection{The Star-formation (Kennicutt-Schmidth) Law}
\label{sec:KS}
\begin{figure}[!t]
  \centering
  \includegraphics[width=0.5\textwidth]{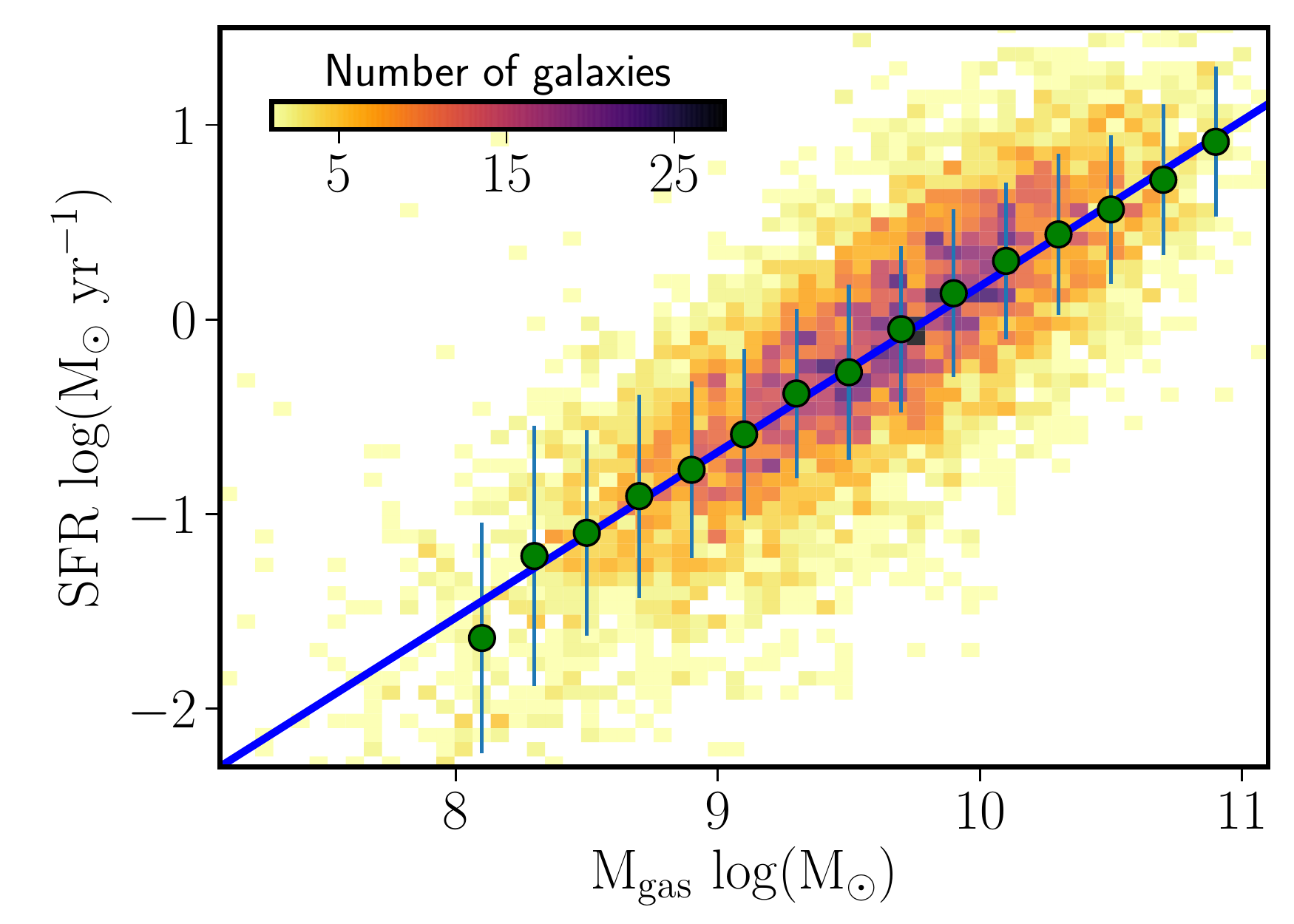}
  \caption{ The integrated Star-formation law for 5600 galaxies included in the MaNGA survey. Both parants are derived using optical observables (see Sec.~\ref{sec:KS} for details). The blue line is a ODR fitting of the median values of SFR at different bins of $\mathrm{M_{gas}}$ (green circles). The slope of this line match the slope using different observables for SFR and $\mathrm{M_{gas}}$  \citep{Gao_2004,Kennicutt_2012}.  }
  \label{fig:KS}
\end{figure}

The star-formation law (also referred as the Kennicutt-Schmidt law, KS-law) correlates the star formation rate (SFR) with the amount of cold gas. For a detailed review on the topic, the reader is refereed to \cite{Kennicutt_2012}. 
Projected in the sky, this relation is measured in surface densities (\mbox{$\Sigma_{\rm SFR} = \Sigma^{N}_{\rm gas}$}), depending on the scales, different components of the cold gas are measured (i.e., neutral, molecular, and dense), and the proxy for SFR the index  $N$ varies \citep{Kennicutt_2012}. This law is also presented in terms of extensive terms.  

In Fig.~\ref{fig:KS}, we plot the relation between the integrated SFR and the molecular gas ($\mathrm{M_{gas}}$) for a sample of 5600 galaxies included in the latest internal MaNGA data release where it was possible to determine $\mathrm{M_{gas}}$. SFR is derived using the dust-corrected luminosity of the H$\alpha$ line, whereas the gas mass ($\mathrm{M_{gas}}$) is obtained using as calibrator the optical extension ($\mathrm{A_v}$) from  the H$\alpha$/H$\beta$ Balmer decrement, following \citep{Barrera-Ballesteros_2019}. The two-dimensional histogram shows the trend of these two parameters with SFR increasing for galaxies with large $\mathrm{M_{gas}}$. The green circles represent the median SFR for different bins of $\mathrm{M_{gas}}$, their errorbars represent the standard deviation of the SFR at each of those bins. The blue line shows the best Orthogonal Distance Regression (ODR) fit of a linear fit to these median values. The best relation is described as:
\begin{equation}
    \log(\mathrm{SFR}) = a + b \times \log(\mathrm{M_{gas}})
\end{equation}

with \mbox{$a = (-8.34 \pm 0.16)\, \mathrm{M_{\odot}\,\,yr^{-1}}$} and $b = (0.85 \pm 0.1)  \,\mathrm{yr^{-1}}$. We note that the slope of this relation is in very good agreement with the one derived using direct observations of the cold dense gas and SFR with IR observables in spiral galaxies \citep[e.g.,][]{Gao_2004, Kennicutt_2012}. This strong correlation suggests the deeply interplay between the amount of gas available to produce new generations of stars and the amount of those stars. As we will see below, as these relations holds at kpc scales, the physical interpretation has to be related to process independent of the scale.   

\subsection{The Star-Formation Main Sequence (SFMS)}
\label{sec:SFMS}
\begin{figure}[!t]
\centering
\includegraphics[width=0.5\textwidth]{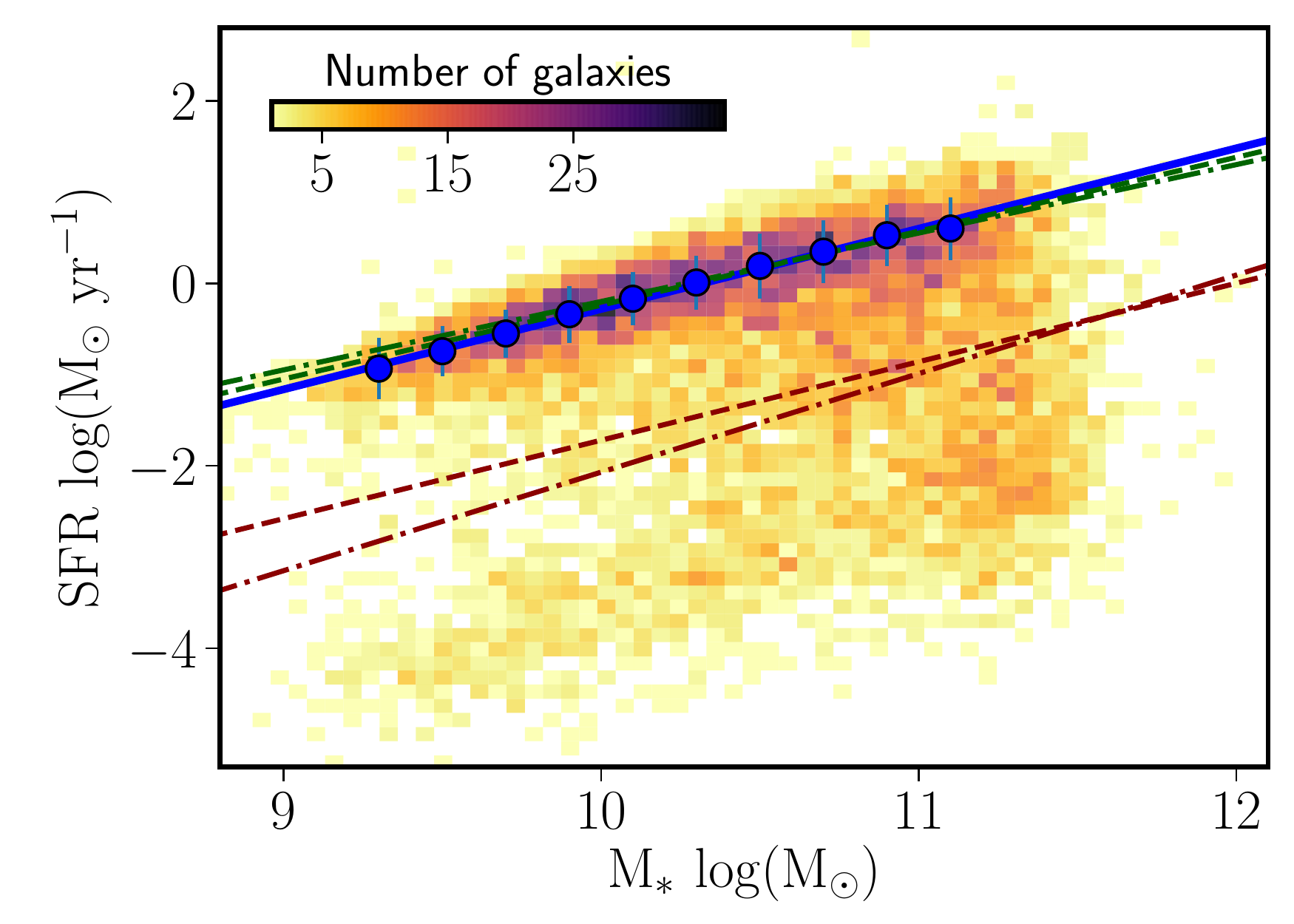}
\caption{Integrated SFR against the total stellar mass ($\mathrm{M_{\ast}}$) for 8000 galaxies included in the MaNGA survey. The blue line is a ODR fitting of the median values of SFR at different bins of $\mathrm{M_{\ast}}$ (green circles). To perform the ODR fitting, we select those galaxies with H$\alpha$ equivalent width EW(H$\alpha$) $>$ 10 \AA. The slope of this line match the trend derived for different studies using star-forming galaxies observed within different IFS surveys. Green and red dashed-line show the SFMS and RGS derived for the CALIFA survey \citep{Cano-Diaz_2016}.Green and red dot-dashed lines represent the best-fit derived using the MaNGA survey \citep{Cano-Diaz_2019}.}
  \label{fig:SFMS}
\end{figure}
With large data set from spectroscopic surveys it was clear the analogous of the CMD. In Fig.\ref{fig:SFMS} we plot the relation between the SFR and the total stellar mass ($\mathrm{M_{\ast}}$) for 8004 MaNGA galaxies. As the color in the CMD, different authors have notice that there is a clear bimodality on the SFR \citep[e.g., ][]{Kauffmann_2003, Baldry_2004, Brinchmann_2004}. Galaxies with high SFR are well described by a tight correlation (known as the Star-formation Main Sequence, SFMS); as $\mathrm{M_{\ast}}$ increases the population of galaxies is divided between galaxies located in the SFMS and those with much less SFR for the same $\mathrm{M_{\ast}}$. Galaxies located in this region of the SFR-$\mathrm{M_{\ast}}$ distribution are considered as part of the 'Retired Galaxies  Sequence' \citep[RGS, e.g., ]{Cano-Diaz_2016}.

Different authors have notice that an excellent parameter to  differentiate galaxies from the SFMS and the RGS is the equivalent width of the H$\alpha$ emission line \citep[EW(H$\alpha$), e.g., ][]{Cid-Fernandes_2010, Sanchez_2013}. Galaxies/regions with large values of EW(H$\alpha$) are actively star-forming. To derive the best fit of the SFMS we select those galaxies with EW(H$\alpha$) $>$ 10 \AA\ measured at their effective radius (3903 galaxies).  In Fig.~\ref{fig:KS} we show the best ODR fit to the SFMS for the MaNGA sample. The best fit is in very good agreement with previous estimations of the SFMS using different IFS samples (see reference in Fig.\ref{fig:SFMS}). Since the SFR-$\mathrm{M_{\ast}}$ plane compares the recent star formation with the integral of the star-formation history (SFH) across cosmic times, the existence of a bimodality can be considered as a record of two different channels of galaxy evolution. Massive galaxies tend to exhibit a sharp SFH in comparison to low-mass galaxies, this effect is known as downsizing \citep{Thomas_2005,Cimatti_2006,Fontanot_2009}. Similar to the CMD \citep{Schawinski_2014}, the location of a galaxy in the   SFR-$\mathrm{M_{\ast}}$ plane is strongly dependent of the morphology \citep[e.g., ][]{Bluck_2019}. 

\subsection{The Mass-Metallicity Relation (MZR)}
\label{sec:MZR}
\begin{figure}[!t]
\centering
\includegraphics[width=0.5\textwidth]{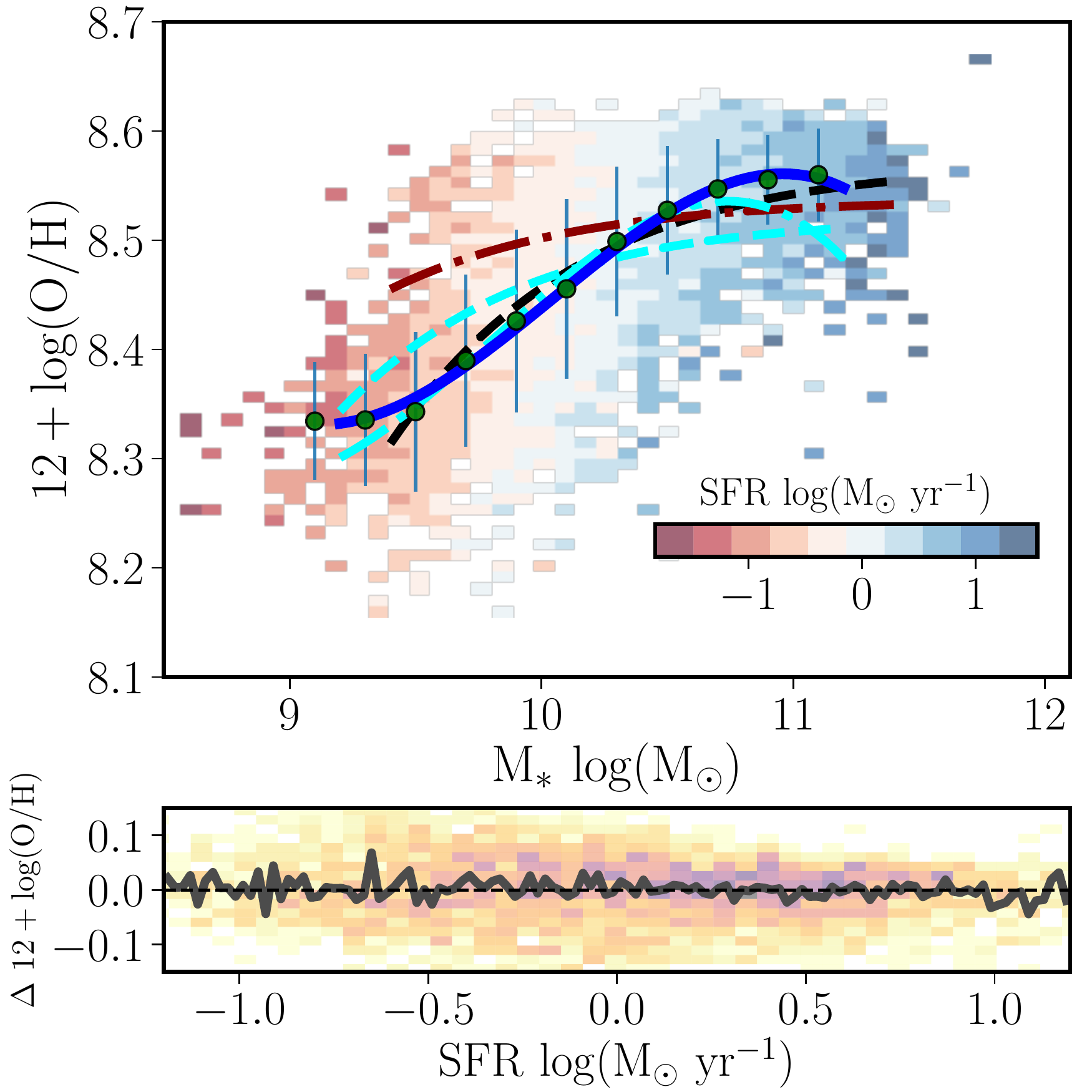}
\caption{The Mass-Metallicity relation (MZR) for 3310 galaxies included in the MaNGA survey is shown in the top panel. The relation is colored by the SFR of the galaxies. Note the almost vertical lines in color for a given SFR bin. The blue line is a forth-degree polynomial fitting of the median values of metallicity at different bins of $\mathrm{M_{\ast}}$ (green circles). The black dashed, red dot-dashed, cyan dashed, and cyan dot-dashed lines represent descriptions of the MZR for previous studies using IFS datasets \citep[MaNGA, CALIFA, and SAMI; ][respectively]{Barrera-Ballesteros_2017, Sanchez_2017, Sanchez_2019}. The bottom panel shows the distribution of the residuals of the MZR against SFR. The black line represents the median residual. No significant trend is found between the residuals of the MZR and the SFR. }
  \label{fig:MZR}
\end{figure}
Four decades ago, it was evident the close relation between the ionized gas metallicity derived from emission lines in H{\sc II} regions/galaxies and the total stellar mass \citep[e.g., ][]{Lequeux_1979, Skillman_1989, Zaritsky_1994}. However,  in the seminal study by \cite{Tremonti_2004}, using single-fiber spectroscopy for more than 53000 star-forming galaxies included in the SDSS survey, they found a very tight correlation (scatter $\sim$ 0.1 dex) between the metallicity (12+$\log({\rm O/H})$) and $\mathrm{M_{\ast}}$ covering several orders of magnitude in stellar mass.

In  Fig.~\ref{fig:MZR} we plot the MZR relation for 3310 star-forming galaxies selected previously in Sec.\ref{sec:SFMS}. The trend of the metallicity observed here is very similar as the one studied by \cite{Tremonti_2004}. As $\mathrm{M_{\ast}}$ increases so does the metallicity, reaching a plateau at massive galaxies ( $\log(\mathrm{M_{\ast}/M{\odot}}) \sim$ 10.5). Following \cite{Barrera-Ballesteros_2017}, we measure the metallicity at the effective radius of the selected MaNGA galaxies. To estimate the metallicity we use the strong-line calibrator derived by \cite{Marino_2013}. We also overplot different estimations of the MZR following a homogeneous procedure as the one used here for different IFS surveys (see caption in Fig.~\ref{fig:MZR}).
This comparison highlight the robustness of the MZR despite the sample of selected galaxies. 

Using the same SDSS it has also been proposed a secondary correlation between the SFR and the MZR \cite{Ellison_2008}. Later different authors called this relation a fundamental metallicity relation or plane \citep[FMR, ][]{Mannucci_2010, Lara-Lopez_2010, Maiolino_2019}. In this scenario, the scatter of the MZR is driven by the SFR. For a given stellar mass, galaxies with large SFR tend to have low metallicities in comparison to galaxies with low SFR. In Fig.~\ref{sec:MZR}, we color the galaxies in the MZR for different SFR bins. We note that for a given $\mathrm{M_{\ast}}$ the SFR color is rather constant for different metallicities. To further quantify the possible impact of the SFR in the MZR we plot in the bottom panel of  Fig.~\ref{sec:MZR} its residual (with respect to the best forth-degree polynomial fit, blue line) against the SFR. We find that the median value of the scatter does not strongly correlate with the SFR. In other words, this implies that introducing the SFR as a secondary parameter does not reduce the observed scatter of the MZR. This lack or small impact of the SFR in the MZR has been observed in different works \citep[e.g., ][]{Kashino_2016, Telford_2016}. In particular, we have explored this dependence (or lack of thereof) for galaxies within all large IFS surveys using a heterogenous set of calibrators \citep[MaNGA, CALIFA, and SAMI;][ respectively]{Barrera-Ballesteros_2017, Sanchez_2017, Sanchez_2019}. We find little to none relation of the SFR or SSFR to the MZR. Including the SFR as second parameter does not reduce the scatter of the MZR. Our results suggest that rather than the metallicity be regulated by global flows of gas (thus related with recent global SFR), it has a local origin (see Sec.\ref{sec:rMZR}). Possibly the ionized gas metallicity is a record of the local SFH. 
\section{Scaling relations at kpc scales}
\label{sec:Local}
\subsection{The resolved Star-formation Law}
\label{sec:rKS}
\begin{figure}[!t]
\centering
\includegraphics[width=0.5\textwidth]{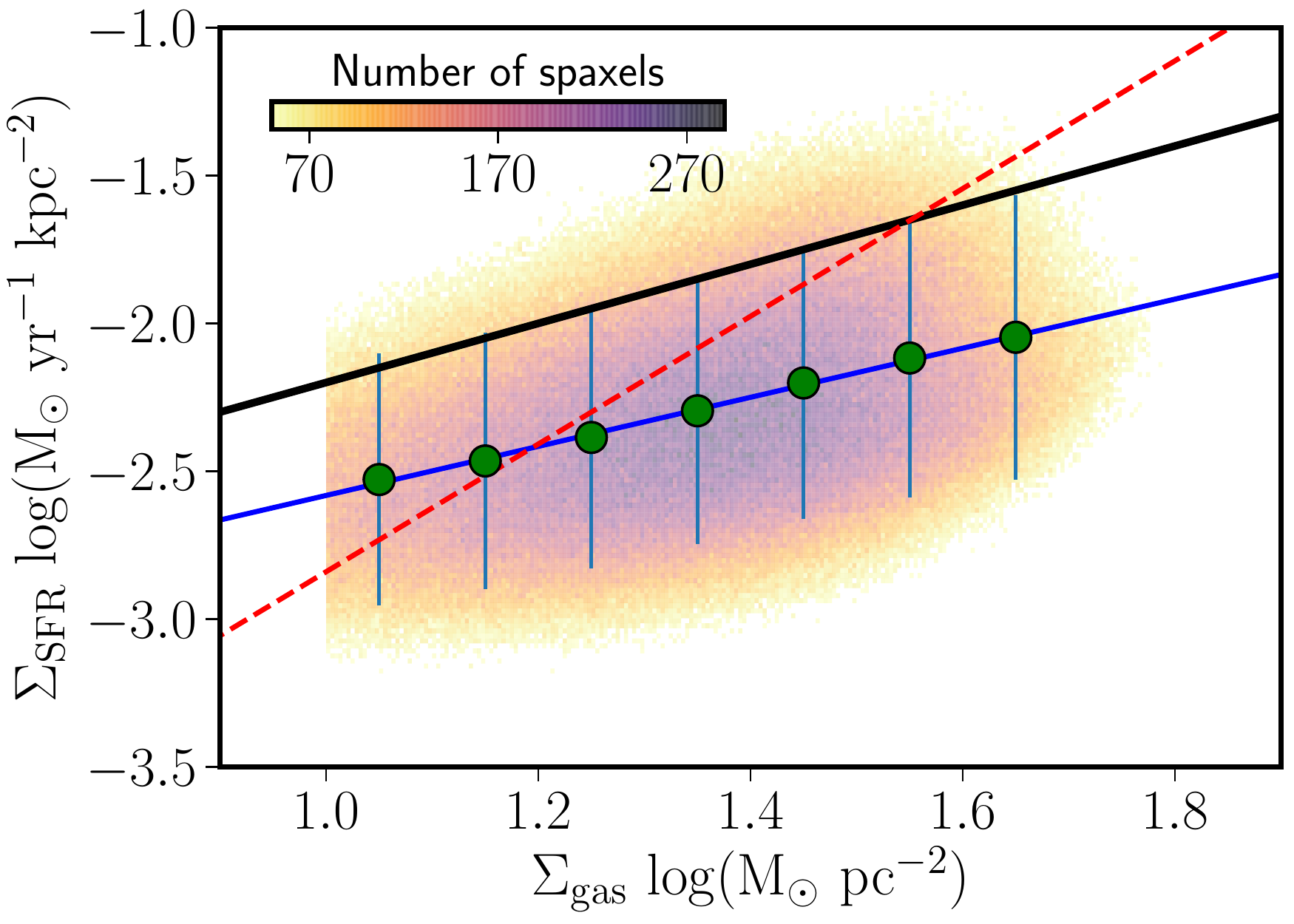}
\caption{ The star-formation law for 2.38$\times\,10^6$ spaxels included in 5382 MaNGA galaxies. $\Sigma_{\rm gas}$ is derived using as proxy the optical extinction (see details in Sec.\ref{sec:KS}). The derived KS-law follows similar trend as those derived using direct observations of the molecular gas \citep[black-dashed line, EDGE survey; ][]{Bolatto_2017}, and those using other IFS surveys \citep[red dot-dahed line, CALIFA survey ; ][]{Barrera-Ballesteros_2019}.}
  \label{fig:rKS}
\end{figure}
With radio telescopes in interferometric mode it has been possible to trace the distribution of the cold gas component at kpc scales ($\Sigma_{\rm gas}$) in nearby galaxies as well as the SFR density ($\Sigma_{\rm SFR}$). This allows to explore the KS-law at (sub-)kpc scales in extragalactic objects \citep[e.g., ][]{Bigiel_2008,Leroy_2008,Kennicutt_2012}. In Fig.~\ref{fig:rKS} we plot this relation for more than  2.38$\times\,10^6$ regions (or spaxels). As in Sec.~\ref{sec:KS}, we use the ${\rm A_V}$ as proxy for $\Sigma_{\rm gas}$.  Similar to the extensive properties, $\Sigma_{\rm SFR}$ tightly correlates with $\Sigma_{\rm gas}$. The KS-law derived here is similar as those derived using direct observations of the molecular gas \citep{Bolatto_2017}. 

As mentioned in the review by \cite{Kennicutt_2012}, the existence of a kpc and global KS-law in extragalactic targets maybe just due to averaging in large scales of physical processes occurring at scales of pcs. Even more as they argue, there are two possible, maybe competing, scenarios that explain the scaling relations observed at pc scales in the Milky Way. In the first scenario star-formation is locally controlled within molecular clouds. Therefore, the properties and amount of gas in the molecular clouds control the efficiency to form new stars \citep[e.g., ][]{Krumholz_2005}. In the other scenario, the star-formation is controlled mainly by galactic-scales processes, in particular dynamic ones \citep[e.g., ][]{Ostriker_2010}. 
\subsection{The Resolved Star-formation and Retired Sequences}
\label{sec:rSFMs}
\begin{figure}[!t]
\centering
\includegraphics[width=0.5\textwidth]{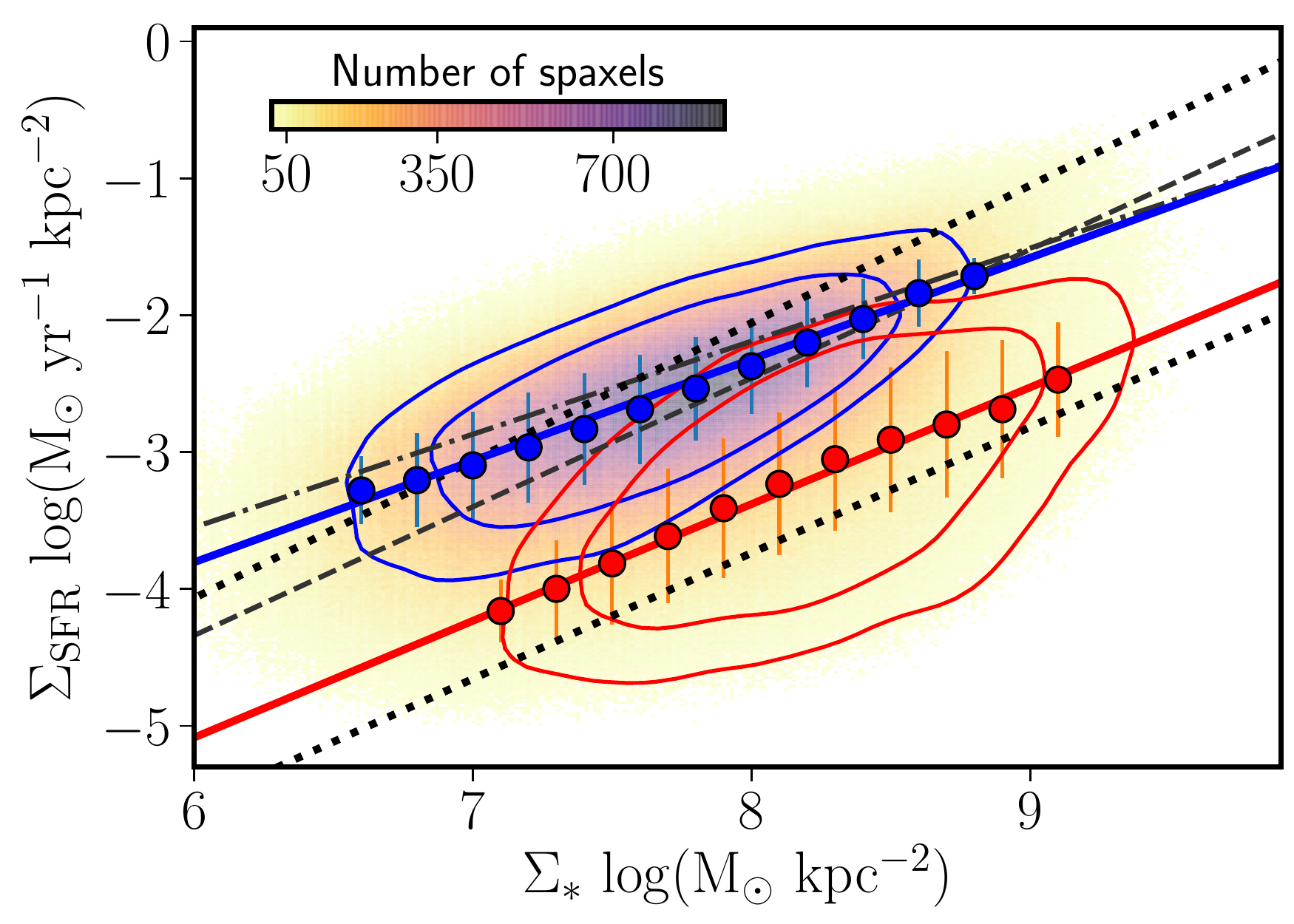}
\caption{$\Sigma_{\rm SFR}$ against $\Sigma_{\ast}$ for 6.06 $\times\,10^6$ spaxels located in 7687 galaxies included in the MaNGA survey. The distribution is color-codded according to the density of spaxels. Each of the blue and red contours enclose 80\% and 60\% of the so called resolved star-formation (rSFS) and resolved retired sequences (rRS), respectively (see details in the text). Blue and red lines represent the best fits of the medians  of $\Sigma_{\rm SFR}$ for bins of  $\Sigma_{\ast}$  for the rSFS and rRS, respectively. We overplot the estimation of the rSFS from other IFS estudies \citep[dot-dashed, dashed and dotted lines represent the fits from the CALIFA, and MaNGA estudies;][respectively]{Cano-Diaz_2016,Cano-Diaz_2019, Hsieh_2017} as well as the estimation of the rRS \citep[lower dotted line][]{Hsieh_2017}. As for the integrated properties (see  Fig.~\ref{fig:SFMS}), the bimodality is also present at kpc scales.  
}
\label{fig:rSFMS}
\end{figure}
Similar to what we observe above in the SFR-$\mathrm{M_{\ast}}$ plane (see Sec.~\ref{sec:SFMS}), there is a clear bimodality observed in the SFR surface density ($\Sigma_{\rm SFR}$) plotted against the stellar mass density ($\Sigma_{\ast}$) at kpc scales in the nearby Universe (see Fig.~\ref{fig:rSFMS}). As for the SFMS and RGS, a good proxy to differentiate the resolved star-formation sequence (rSFS) and the retire sequence (rRS) is via the EW(H$\alpha$). Following \cite{Cano-Diaz_2016}, we select spaxels with EW(H$\alpha$) $>$ 6 \AA\  as part of the rSFS; spaxels with smaller  EW(H$\alpha$) are considered as rRS spaxels. The best fit of both relations are in very good agreement with other studies, even using different IFS surveys (see caption of Fig.~\ref{fig:rSFMS} for details). Different studies have suggested also the strong impact of the morphology and spatial distribution of the rSFS and rRS \citep[e.g., ][]{Medling_2018,Cano-Diaz_2019}. These trends suggest that the processes responsible for controlling SF are of a local nature (self-regulation, local outflows) whereas quenching processes may be due to galaxy-scale processes \citep{Sanchez_2019_rev}.

\subsection{The Resolved Mass Metallicity relation}
\label{sec:rMZR}
\begin{figure}[!t]
\centering
\includegraphics[width=0.5\textwidth]{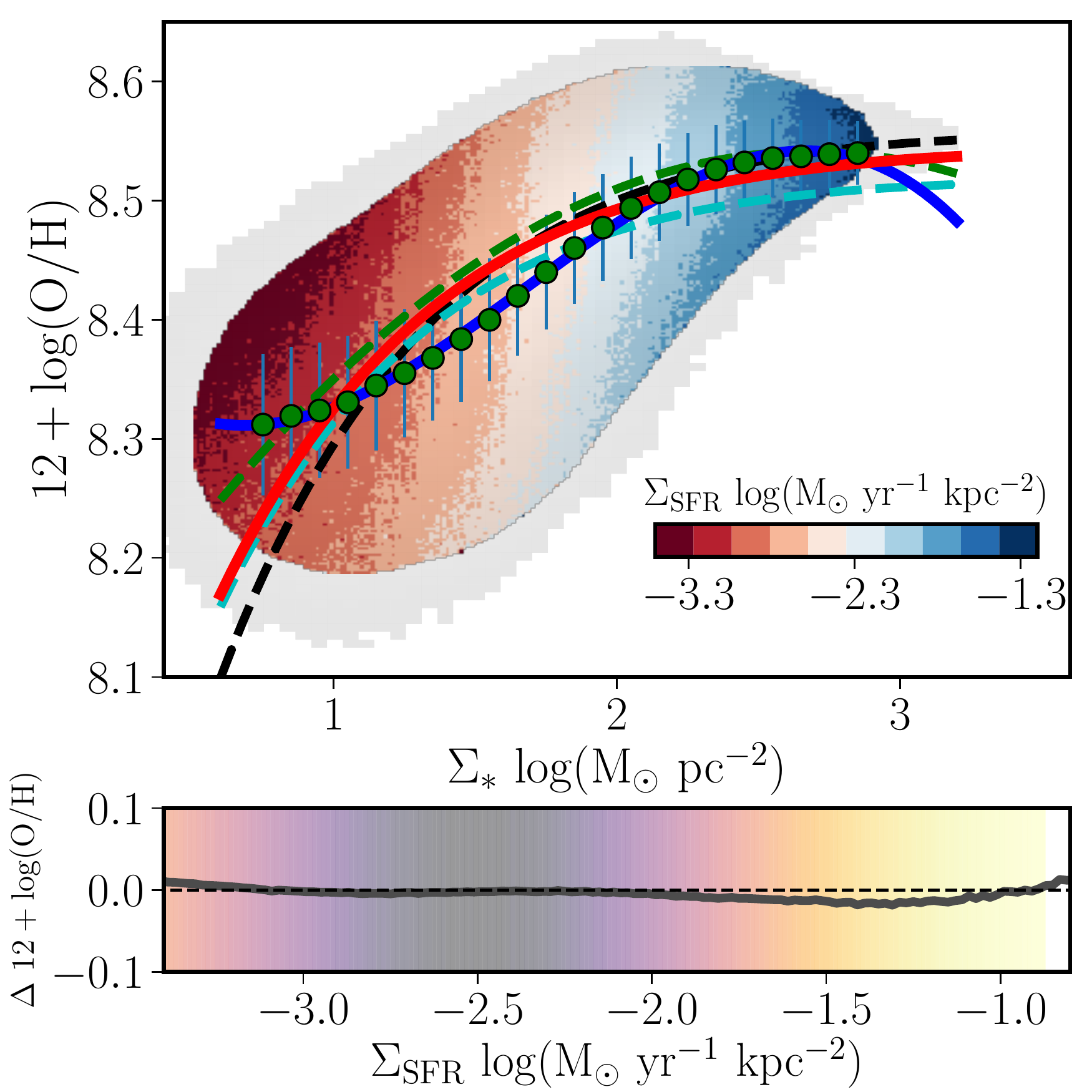}
\caption{\textit{Top pannel} The spatially-resolved mass-metallicity relation (rMZR) for 3.59 $\times\,10^6$ spaxels located in 4904 galaxies included in the MaNGA survey. The 80\% of the distribution of the rMZR is color-codded with their respect $\Sigma_{\rm SFR}$. For a given $\Sigma_{\ast}$ bin there is little variation of $\Sigma_{\rm SFR}$ in the rMZR plane. The blue and red lines represent different fitted functions commonly used. The fit is done for the median metallicities (green circles). The green, cyan, and black dashed lines represent the best fit derived in different IFS surveys \citep[PINGS, CALIFA, MaNGA;][ respectively]{Rosales-Ortega_2012, Sanchez_2013, Barrera-Ballesteros_2016}. \textit{Bottom pannel} The residuals of the rMZR (with respect to the blue solid line) against the $\Sigma_{\rm SFR}$. As for the MZR, the median residuals shows a flat trend, indicating that the $\Sigma_{\rm SFR}$ does not reduce the scatter of the rMZR.     
}
\label{fig:rMZR}
\end{figure}
In top panel of Fig.\ref{fig:rMZR} we plot the relation of the local metallicity -- using the same calibrator as in Sec.\ref{sec:MZR} --  and $\Sigma_{\ast}$ (rMZR) for the sample of \textit{bona fide} star forming regions (i.e., spaxels with line ratios below the \cite{Kauffmann_2003_BPT} demarcation line and EW(H$\alpha$) $>$ 10\AA). It is evident the similarity in shape with the global MZR (see Sec.\ref{sec:MZR}). In fact, the MZR can be recovered by integrating from the rMZR \citep{Rosales-Ortega_2012, Barrera-Ballesteros_2016}. Using a exponential function (red-solid line), the best fit to the median values of the rMZR is in very good agreement with previous estimations of the rMZR (see caption of Fig.~\ref{fig:rMZR} for details). A forth-degree polynomial function was also fitted to the data (solid blue line) yielding a more tight fit to the median values. The rMZR has been found to be fairly independent of the total stellar mass -- except for the less massive galaxies -- and it does explain fairly well the observed metallicity gradients in star-forming galaxies \citep{Barrera-Ballesteros_2016}. 

Recently, different authors have explore the possible secondary dependence of the metallicity with respect to $\Sigma_{\rm SFR}$ \citep[e.g., ]{Maiolino_2019,Sanchez-Menguiano_2019}. In other words, a resolved FMR. In contrast to these studies, we find that for a given $\Sigma_{\ast}$, $\Sigma_{\rm SFR}$ is rather constant (see color distribution in Fig.~\ref{fig:rMZR}). We further quantify this possible relation by plotting the scatter of the relation with respect the blue solid line. We find that the residuals of the rMZR does not correlate with $\Sigma_{\rm SFR}$. Using the exponential fit to derive the residual may yield a mild positive relation of the scatter only for spaxels with very low  $\Sigma_{\rm SFR}$ 
($<$ -3.0 $\log(\mathrm{M_{\odot}\,\,yr^{-1}\,\,kpc^{-2}})$). These results suggest that as for the integrated relation there is no clear evidence of a resolved FMR or a secondary relation of the rMZR  with $\Sigma_{\rm SFR}$. The existence of a local counterpart of the MZR (largely independent of the morphology and the stellar mass) suggests that the physical process(es) responsible for this scaling relation has a local origin. Even more, the lack of relation with the $\Sigma_{\rm SFR}$ indicates that other parameters may play an important role shapping the local abundance in star-forming galaxies \citep[e.g., gas fraction, mass-loading factor; ][]{Barrera-Ballesteros_2018}.
%
%
%
%
\begin{acknowledgement}
The author is grateful with the LOC/SOC of the XVI LARIM, without their support it would not been possible for him to assist. The author thanks S.F. S\'anchez for providing the data analysis products of the MaNGA datacubes to present the plots in this proceeding. The author acknowledges support from the CONACYT grant CB-285080 and FC-2016-01-1916, and funding from the PAPIIT-DGAPA-IA101217 and PAPIIT-DGAPA-IA100420 (UNAM) projects. Funding for the Sloan Digital Sky Survey IV has been provided by the Alfred P. Sloan Foundation, the U.S. Department of Energy Office of Science, and the Participating Institutions. Funding for the Sloan Digital Sky Survey IV has been provided by the Alfred P. Sloan Foundation, the U.S. Department of Energy Office of Science, and the Participating Institutions. SDSS-IV acknowledges
support and resources from the Center for High-Performance Computing at the University of Utah. The SDSS web site is www.sdss.org.
SDSS-IV is managed by the Astrophysical Research Consortium for the Participating Institutions of the SDSS Collaboration including the Brazilian Participation Group, the Carnegie Institution for Science, Carnegie Mellon University, the Chilean Participation Group, the French Participation Group, Harvard-Smithsonian Center for Astrophysics, Instituto de Astrof\'isica de Canarias, The Johns Hopkins University, Kavli Institute for the Physics and Mathematics of the Universe (IPMU) / University of Tokyo, the Korean Participation Group, Lawrence Berkeley National Laboratory, 
Leibniz Institut f\"ur Astrophysik Potsdam (AIP),  Max-Planck-Institut f\"ur Astronomie (MPIA Heidelberg), 
Max-Planck-Institut f\"ur Astrophysik (MPA Garching), Max-Planck-Institut f\"ur Extraterrestrische Physik (MPE), 
National Astronomical Observatories of China, New Mexico State University, 
New York University, University of Notre Dame, 
Observat\'ario Nacional / MCTI, The Ohio State University, 
Pennsylvania State University, Shanghai Astronomical Observatory, 
United Kingdom Participation Group,
Universidad Nacional Aut\'onoma de M\'exico, University of Arizona, 
University of Colorado Boulder, University of Oxford, University of Portsmouth, 
University of Utah, University of Virginia, University of Washington, University of Wisconsin, 
Vanderbilt University, and Yale University.
\end{acknowledgement}


\bibliographystyle{baaa}
\small
\bibliography{main}

\begin{thebibliography}{45}
\providecommand{\natexlab}[1]{#1}

\bibitem[{{Baldry} et~al.(2004)}]{Baldry_2004}
{Baldry} I.K., et~al., 2004, \apj, 600, 681

\bibitem[{{Barrera-Ballesteros} et~al.(2016)}]{Barrera-Ballesteros_2016}
{Barrera-Ballesteros} J.K., et~al., 2016, \mnras, 463, 2513

\bibitem[{{Barrera-Ballesteros} et~al.(2017)}]{Barrera-Ballesteros_2017}
{Barrera-Ballesteros} J.K., et~al., 2017, \apj, 844, 80

\bibitem[{{Barrera-Ballesteros} et~al.(2018)}]{Barrera-Ballesteros_2018}
{Barrera-Ballesteros} J.K., et~al., 2018, \apj, 852, 74

\bibitem[{{Barrera-Ballesteros} et~al.(2019)}]{Barrera-Ballesteros_2019}
{Barrera-Ballesteros} J.K., et~al., 2019, arXiv e-prints, arXiv:1911.09677

\bibitem[{{Bigiel} et~al.(2008)}]{Bigiel_2008}
{Bigiel} F., et~al., 2008, \aj, 136, 2846

\bibitem[{{Bluck} et~al.(2019)}]{Bluck_2019}
{Bluck} A.F.L., et~al., 2019, \mnras, 2839

\bibitem[{{Bolatto} et~al.(2017)}]{Bolatto_2017}
{Bolatto} A.D., et~al., 2017, \apj, 846, 159

\bibitem[{{Brinchmann} et~al.(2004)}]{Brinchmann_2004}
{Brinchmann} J., et~al., 2004, \mnras, 351, 1151

\bibitem[{{Bundy} et~al.(2015)}]{Bundy_2015}
{Bundy} K., et~al., 2015, \apj, 798, 7

\bibitem[{{Cano-D{\'\i}az} et~al.(2016)}]{Cano-Diaz_2016}
{Cano-D{\'\i}az} M., et~al., 2016, \apjl, 821, L26

\bibitem[{{Cano-D{\'\i}az} et~al.(2019)}]{Cano-Diaz_2019}
{Cano-D{\'\i}az} M., et~al., 2019, \mnras, 488, 3929

\bibitem[{{Cid Fernandes} et~al.(2010)}]{Cid-Fernandes_2010}
{Cid Fernandes} R., et~al., 2010, \mnras, 403, 1036

\bibitem[{{Cimatti} et~al.(2006){Cimatti}, {Daddi} \& {Renzini}}]{Cimatti_2006}
{Cimatti} A., {Daddi} E., {Renzini} A., 2006, \aap, 453, L29

\bibitem[{{Ellison} et~al.(2008)}]{Ellison_2008}
{Ellison} S.L., et~al., 2008, \apjl, 672, L107

\bibitem[{{Fontanot} et~al.(2009)}]{Fontanot_2009}
{Fontanot} F., et~al., 2009, \mnras, 397, 1776

\bibitem[{{Gao} \& {Solomon}(2004)}]{Gao_2004}
{Gao} Y., {Solomon} P.M., 2004, \apj, 606, 271

\bibitem[{{Hsieh} et~al.(2017)}]{Hsieh_2017}
{Hsieh} B.C., et~al., 2017, \apjl, 851, L24

\bibitem[{{Hubble}(1926)}]{Hubble_1926}
{Hubble} E.P., 1926, \apj, 64, 321

\bibitem[{{Kashino} et~al.(2016)}]{Kashino_2016}
{Kashino} D., et~al., 2016, \apjl, 823, L24

\bibitem[{{Kauffmann} et~al.(2003{\natexlab{a}})}]{Kauffmann_2003}
{Kauffmann} G., et~al., 2003{\natexlab{a}}, \mnras, 341, 54

\bibitem[{{Kauffmann} et~al.(2003{\natexlab{b}})}]{Kauffmann_2003_BPT}
{Kauffmann} G., et~al., 2003{\natexlab{b}}, \mnras, 346, 1055

\bibitem[{{Kennicutt} \& {Evans}(2012)}]{Kennicutt_2012}
{Kennicutt} R.C., {Evans} N.J., 2012, \araa, 50, 531

\bibitem[{{Krumholz} \& {McKee}(2005)}]{Krumholz_2005}
{Krumholz} M.R., {McKee} C.F., 2005, \apj, 630, 250

\bibitem[{{Lara-L{\'o}pez} et~al.(2010)}]{Lara-Lopez_2010}
{Lara-L{\'o}pez} M.A., et~al., 2010, \aap, 521, L53

\bibitem[{{Lequeux} et~al.(1979)}]{Lequeux_1979}
{Lequeux} J., et~al., 1979, \aap, 500, 145

\bibitem[{{Leroy} et~al.(2008)}]{Leroy_2008}
{Leroy} A.K., et~al., 2008, \aj, 136, 2782

\bibitem[{{Maiolino} \& {Mannucci}(2019)}]{Maiolino_2019}
{Maiolino} R., {Mannucci} F., 2019, \aapr, 27, 3

\bibitem[{{Mannucci} et~al.(2010)}]{Mannucci_2010}
{Mannucci} F., et~al., 2010, \mnras, 408, 2115

\bibitem[{{Marino} et~al.(2013)}]{Marino_2013}
{Marino} R.A., et~al., 2013, \aap, 559, A114

\bibitem[{{Medling} et~al.(2018)}]{Medling_2018}
{Medling} A.M., et~al., 2018, \mnras, 475, 5194

\bibitem[{{Ostriker} et~al.(2010){Ostriker}, {McKee} \&
  {Leroy}}]{Ostriker_2010}
{Ostriker} E.C., {McKee} C.F., {Leroy} A.K., 2010, \apj, 721, 975

\bibitem[{{Rosales-Ortega} et~al.(2012)}]{Rosales-Ortega_2012}
{Rosales-Ortega} F.F., et~al., 2012, \apjl, 756, L31

\bibitem[{{Sanchez}(2019)}]{Sanchez_2019_rev}
{Sanchez} S.F., 2019, arXiv e-prints, arXiv:1911.06925

\bibitem[{{S{\'a}nchez} et~al.(2013)}]{Sanchez_2013}
{S{\'a}nchez} S.F., et~al., 2013, \aap, 554, A58

\bibitem[{{S{\'a}nchez} et~al.(2016)}]{Sanchez_2016}
{S{\'a}nchez} S.F., et~al., 2016, \rmxaa, 52, 171

\bibitem[{{S{\'a}nchez} et~al.(2017)}]{Sanchez_2017}
{S{\'a}nchez} S.F., et~al., 2017, \mnras, 469, 2121

\bibitem[{{S{\'a}nchez} et~al.(2019)}]{Sanchez_2019}
{S{\'a}nchez} S.F., et~al., 2019, \mnras, 484, 3042

\bibitem[{{S{\'a}nchez-Menguiano} et~al.(2019)}]{Sanchez-Menguiano_2019}
{S{\'a}nchez-Menguiano} L., et~al., 2019, \apj, 882, 9

\bibitem[{{Schawinski} et~al.(2014)}]{Schawinski_2014}
{Schawinski} K., et~al., 2014, \mnras, 440, 889

\bibitem[{{Skillman} et~al.(1989){Skillman}, {Kennicutt} \&
  {Hodge}}]{Skillman_1989}
{Skillman} E.D., {Kennicutt} R.C., {Hodge} P.W., 1989, \apj, 347, 875

\bibitem[{{Telford} et~al.(2016)}]{Telford_2016}
{Telford} O.G., et~al., 2016, \apj, 827, 35

\bibitem[{{Thomas} et~al.(2005)}]{Thomas_2005}
{Thomas} D., et~al., 2005, \apj, 621, 673

\bibitem[{{Tremonti} et~al.(2004)}]{Tremonti_2004}
{Tremonti} C.A., et~al., 2004, \apj, 613, 898

\bibitem[{{Zaritsky} et~al.(1994){Zaritsky}, {Kennicutt} \&
  {Huchra}}]{Zaritsky_1994}
{Zaritsky} D., {Kennicutt} Robert~C. J., {Huchra} J.P., 1994, \apj, 420, 87

\end{thebibliography}
 
\end{document}